\documentclass[twoside,slac_one]{revtex4}
\usepackage{graphicx}
\usepackage{fancyhdr}
\usepackage{amsmath} % American Mathematics Society standards
\usepackage{bm}% bold math
\usepackage{amsxtra}
\usepackage{amssymb}
\usepackage{amsthm}
\usepackage{latexsym}
\usepackage{lscape}

\begin{document}

\title{Hadron Spectroscopy} 


\author{Adam P. Szczepaniak}
\affiliation{Department of Physics and CEEM, Indiana University, Bloomington, IN 47405, USA} 

\begin{abstract}
Advances in experiment, theory and phenomenology of hadron spectroscopy are discussed. 
We focus on specific developments in the meson sector. 
\end{abstract}

\maketitle

\thispagestyle{fancy}


%%%%%%%%%%%%%%%%%%%%%%%%%%%%%%%%%%
\section{Introduction}

Hadron spectroscopy has played an important role in developing  phenomenology and  gaining insights  into QCD in the non-perturbative domain.  In the days of the $S$-matrix bootstrap 
 it was assumed that hadrons originate  from interactions between  
  mesons and baryons.  $S$-matrix analyticity together  with requirements of crossing-symmetry and unitarity were supposed to provide constraints that are sufficient  for computing 
    the scattering amplitude. With  Regge theory determining the asymptotic behavior,  dispersion techniques were used to generate dynamical equations for the scattering amplitude.  While  the bootstrap program  did enjoy some success \cite{Collins:1971ff}
  these  general requirements  from the $S$-matrix theory were  found to be  insufficient to unambiguously determine hadronic amplitudes. Nowadays these ambiguities can be traced to existence 
    of QCD bound states,  however even after 40 years of hadron physics we are still far from 
     being able to construct hadronic amplitudes from first principles. 
     
       The symmetry pattern of  the hadron spectrum lead to   development of the quark model  which together with the deep inelastic phenomena are the  pillars of QCD.  
         Once QCD has been established as the underlying, microscopic theory of strong interactions 
          interest in hadron physics   shifted towards  ab initio calculation.  
  Even though    some of the amplitude  ambiguities    can be eliminated with the help of 
   QCD {\it e.g.} by relying on the lattice QCD spectrum~\cite{Jozef:2010}  or on the low energy  approximations ~\cite{Pelaez:2003dy}, 
           hadronic reaction theory in the resonance region still largely  relies on phenomenological analysis \cite{Martin:1978}. 
 Thus identification  of new resonances and analysis of  their properties can only be successful if there is  a combined effort involving experimental analysis,  theory and phenomenology. 
 This is particularly pressing now since in recent years new hadron resonances that  seem to escape the established phenomenology have been found in both light and heavy quark sector, and   
  numerous experimental efforts in hadron spectroscopy have been undertaken.

 In this talk I will review the recent experimental, theoretical and phenomenological results with  focus on  the  meson spectrum and indicate the new challenges in the amplitude analysis.

%%%%%%%%%%%%%%%%%%%%%%%%%%%%%%%%%%
\section{The QCD template}
\label{template} 

Hadrons are excited states of the QCD vacuum. Therefore to understand the excitation spectrum one needs first to understand the ground state. In QED, bound states between non-relativistic charges are formed because of the static Coulomb potential and radiation is only weakly coupled. 
In QCD, vacuum fields are  quite different. Vacuum field distribution is expected to have  nontrivial topology with features like  confinement originating from  disorder of  chromo-magnetic domains such as vortices or monopoles~\cite{Greensite:2003bk}. 
 The vacuum fields are also known to affect the non-abelian Coulomb potential and make it (over) confining.  The Wilson string tension between static quarks emerges as the Coulomb system evolves to its  lowest eigenstate and in the process leads to a formation of 
  the chromo-electric  flux tube. The emerging spectrum of hadrons   
     contains the quark model type flavor multiplets  of  the leading Regge trajectories together with daughter trajectories corresponding to radial excitations, with spin-parity  degeneracies  broken by  the hyperfine interactions.  
Since the flux tube is not rigid its excitations could lead  to additional sets of Regge trajectories.  These are termed exotic hadrons. Furthermore, in absence of large strong  van der Waals forces, residual interactions between color-singlet hadrons  are expected to be weak 
 nevertheless they too could produce some residual binding leading to molecular states,  
  in particular for heavier quarks. The best known candidate for a meson-meson molecule is  the $f_0(980)$ resonance in the $K\bar K$ system.

\section{Heavy Quarkonia} 

\subsection{ Charmonium spectrum} 

\begin{figure}[ht]
\centering
\includegraphics[width=150mm]{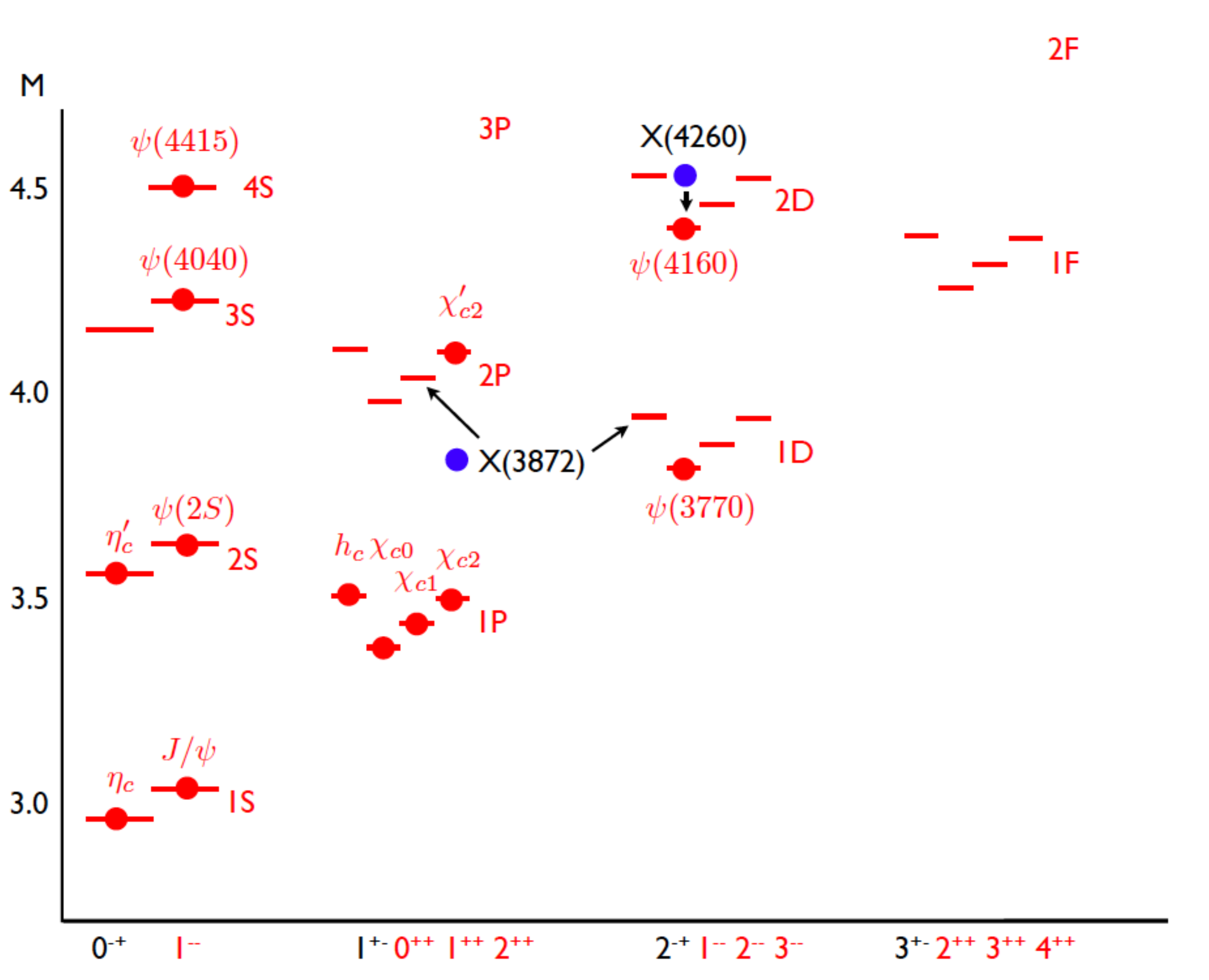}
\caption{Charmonium spectrum. Well established states are shown as red dots and are plotted agains  location of the quark model levels (labeled by the radial quantum number and orbital angular momentum between the $Q{\bar Q}$ pair).  The two well established "exotic" states, the $X(3872)$ and $X(4260)$ are shown as blue dots with arrow indicating possible quark model assignments.  Total spin, $J$, parity $P$ and charge conjugation $C$, $J^{PC}$ quantum numbers are displayed on the horizontal axis.} \label{fig1}
\end{figure}

The established states of charmonium  are schematically shown in Fig.~\ref{fig1} together with the  
   QCD template discussed above. Below the $D{\bar D}$ threshold the quark model predicts $1S$, $2S$
 and $1P$ levels and total quark spin $S=0,1$. A simple potential model predictions based on the Cornell parametrization  of the temporal Wilson loop agree well with the measured spectrum. Above the strong decay threshold the quark model still gives a good description of  the $3S$ and possibly even the  $4S$  levels. Other know states include the two excited vector states, the $\psi(3770)$ and $\psi(4160)$  which can possibly fill in two slots in the $1D$  and $2D$ multiplets, respectively. Also the $\chi'_{c2}$  possibly belonging to the $2P$ multiplet has been identified.

In recent years, Belle, Babar, and CDF have reported signatures of other charmonium-like  states, 
which in literature are refereed to as the  $XYZ$ states. The 2011 edition of the Particle Data Book~\cite{pdg}  lists a total of ten $XYZ's$, however, only two of them have been seen by more then one experiment: the $X(3872)$ and the $Y(4260)$. 

\subsection{$X(3872)$ and meson-molecules } 
The $X(3872)$ was discovered by Belle~\cite{Choi:2003ue} in  exclusive decays  
$B^\pm \to K^\pm \pi^+ \pi^- J/\psi$. Subsequently it 
  was  confirmed by the CDF \cite{Acosta:2003zx} and by now both Belle and Babar measured  several other decay modes. 
 The $X(3872)$ is only $O(MeV)$  away form the  $D\bar D^*$  and $D\bar D\pi$ mass thresholds. 
  The proximity of the $D \bar D^*$ threshold and absence of a possible quark model level nearby    makes the molecular interpretation the most natural,  
   provided the state has $1^{++}$ quantum numbers, {\it i.e} $D\bar D^*$ binds in the $S$-wave. 
  It is worth noting that such a molecule would have size of  several fm's.
   The closet quark model slot available in this mass range is in the $2P$ multiplet which is a 
    few hundred MeV     above the $X(3872)$  while the other potentially available slot,  in the lighter $1D$  multiplet is commonly attributed to the  $\phi(3770)$\cite{Swanson:2006st}. 
    
     Recent Babar measurement of the  angular distribution~\cite{jo}  of $J\psi \omega$ meson pairs 
       from $X(3872)$ decay  favors, however, the  $J^{PC}=2^{-+}$ spin-parity assignment. If this is the case then   the $1D$ quark model assignment would be more likely.  Future data on radiative decays  to other $c\bar c$ states could settle this issue. Both Belle and Babar have so far measured  the radiative decay to $J/\psi$ and $\psi'$ finding rates that are compatible with  those expected from a decay of a compact charmonium   state~\cite{Bhardwaj:2011dj,Aubert:2005rm}.

\begin{figure}[ht]
\centering
\includegraphics[width=150mm]{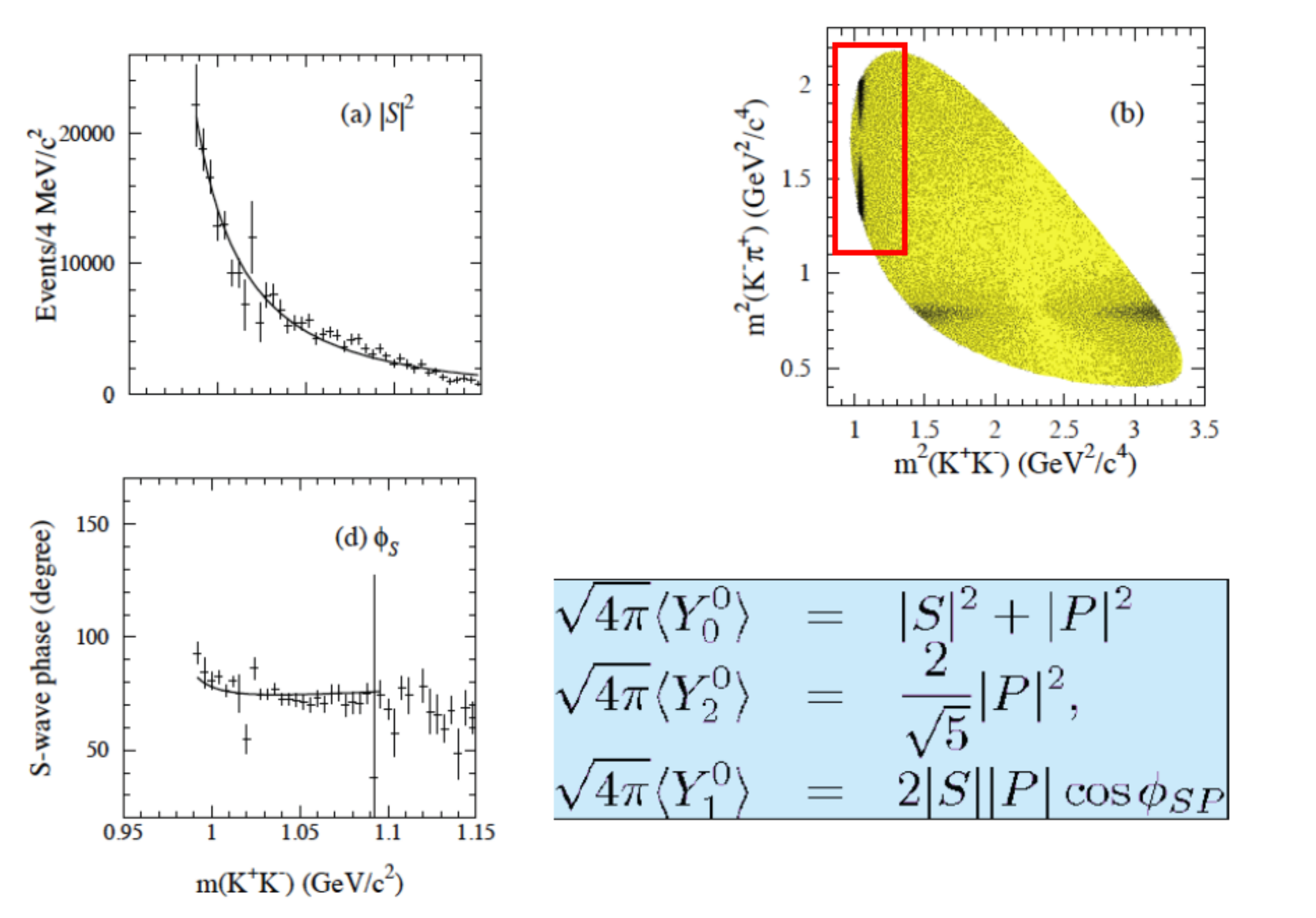}
\caption{Top-Right:  $K^+K^-\pi^+$ Dalitz distribution. The window in the top-left corner shows the $\phi(1020)$ resonance and the clear interference pattern of the $P$ and $S$-wave. Left: Intensity and phase of the $S$ extracted by fitting the $K^+K^-$ angular moment distribution (bottom-right) assuming the $P$-wave $\phi(1020)$ can be described as a 
Breit-Wigner resonance. The solid line represents a fit to a second  Breit-Wigner formula  assuming 
 $S$-wave is dominated by the $f_0(980)$ resonance.}  \label{fig2}
\end{figure}

As  mentioned above a possible molecular state has been known to exist in the $s\bar s$ channel.  
While in the past most  of the data on  light hadron spectrum  came from fixed  target scattering experiments today the light hadron database has been enlarged by data 
  from  heavy flavor decays.  The BaBar collaboration has recently performed Dalitz plot analysis of the $D^+_s \to K^+ K^- \pi^+$ decay\cite{delAmoSanchez:2010yp} ({\it cf. } Fig.~\ref{fig2}). In the $K^+K^-$  mass  window around the $P$-wave $\phi(1020)$ resonance, by exploring the 
   interference between the $S$ and the $P$ wave, they  were able to extract in a model-independent way  the  $K\bar K$ $S$-wave. The $S$-wave intensity and phase was found to be consistent with a resonant  description characteristic of the $f_0(980)$ and it will be interesting to see how these new $S$-wave data impact on the partial wave dispersive analyses~\cite{pelaez-kaminski}

\subsection{$Y(4260)$} 

Besides the $X(3872)$, the $Y(4260)$ is the other $XYZ$ state that has been seen in more then one experiment. Discovered by Babar ~\cite{Aubert:2005rm} in initial state radiation 
$e^+ e^- \to \gamma_{ISR} \pi^+ \pi^- J/\psi$ was confirmed by 
Belle in the same final state  ~\cite{:2007sj}  and  by CLEO ~\cite{He:2006kg}. While there have been attempts to associate the $Y(4260)$ with a quark model state~\cite{felipe},  the closest vector slots in the $4S$ and $2D$ multiplets are more likely to be associated with the $\psi(4415)$ and $\psi(4160)$, respectively. An alternative description as a quark-gluon hybrid will be discussed in Sec.~\ref{hyb} below.

\subsection{ $XYZ$'s in the $3900 -4200$ mass range }
There is strong evidence for other charmonium-like resonances. Unfortunately no single set 
 of resonance parameters consistent with all experimental sightings has so far been found. 
   In 2005  Belle reported     a possible resonance with mass $M=3943\pm 11$ and width  $\Gamma = 87\pm 22$ decaying to $J/\psi \omega$ by analyzing the mass projection of  the Dalitz distribution in  $B\to K\omega J/\psi$ \cite{Abe:2004zs}. Later, a similar resonance was reported  by the same collaboration   in the spectrum of masses recoiling from the $J/\psi$ in the inclusive process $e^+ e^- \to J/\psi + \mbox{ anything}$\cite{Abe:2007jn}. 
 In contrast,  Babar has observed a state decaying to $J\psi \omega$  in the  process
$B^{0,+} \to J/\psi \omega(\pi^+\pi^-\pi^0) K^{0,+}$ with mass $M=3919.1^{+3.8}_{-3.4} \pm 2.0$ and  width,   $\Gamma=31^{+10}_{-8} \pm 5$ \cite{delAmoSanchez:2010jr}, while CDF repots a state with  $M=4143.4^{-2.9}_{-3.0} + \pm 0.6$ and $\Gamma = 15.3^{+10.4}_{-6.1} \pm 2.5$ in $B^{\pm} \to J/\psi \phi K^{\pm}$ ~\cite{Aaltonen:2011at}. Theoretical interpretations based on the valence quark model, multi-quark or meson molecular models have been proposed. More insight  into the production and decay characteristics of these states  is needed before concrete statements pertaining to their  nature can be made.

\subsection{ Charged Charmonia?} 

Belle collaboration performed amplitude analys of the Dalitz plot distribution in $\bar B^0 \to K^- \psi' \pi^+$ and $B^+ \to K^0_s \psi' \pi^+$~\cite{:2009da}. 
In the $K\pi$ channel they included various $K^*$ resonances. While the Dalitz plot does not
  indicate any enhancement in the $\psi' \pi$ channel,  addition of a  Breit-Wigner amplitude corresponding  to a charged charmed resonance clearly improved the quality of the fit. 
Besides resonant amplitude there was also a background term included in fit. 
Even though the  best fit was obtained with the charged $\psi'\pi$ resonance it is should be noted 
  that the background wave was quite significant  and so were the possible kinematical reflections form higher spin {\it e.g.} $K^*_3$ resonances in the $K\pi$ channel. Although $K^*$'s resonances 
   cannot produce sharp peaks when projected onto the $\psi'\pi$ mass distribution,  in general,
    kinematic reflections may be responsible for producing fake signals as statistical fluctuations. An example of a kinematical  effect mimicking a resonance in a crossed channel  was studied in \cite{Dzierba:2003cm}. No charged charmonium states were found in the Babar data~\cite{:2008nk}.

\begin{figure}[ht]
\centering
\includegraphics[width=150mm]{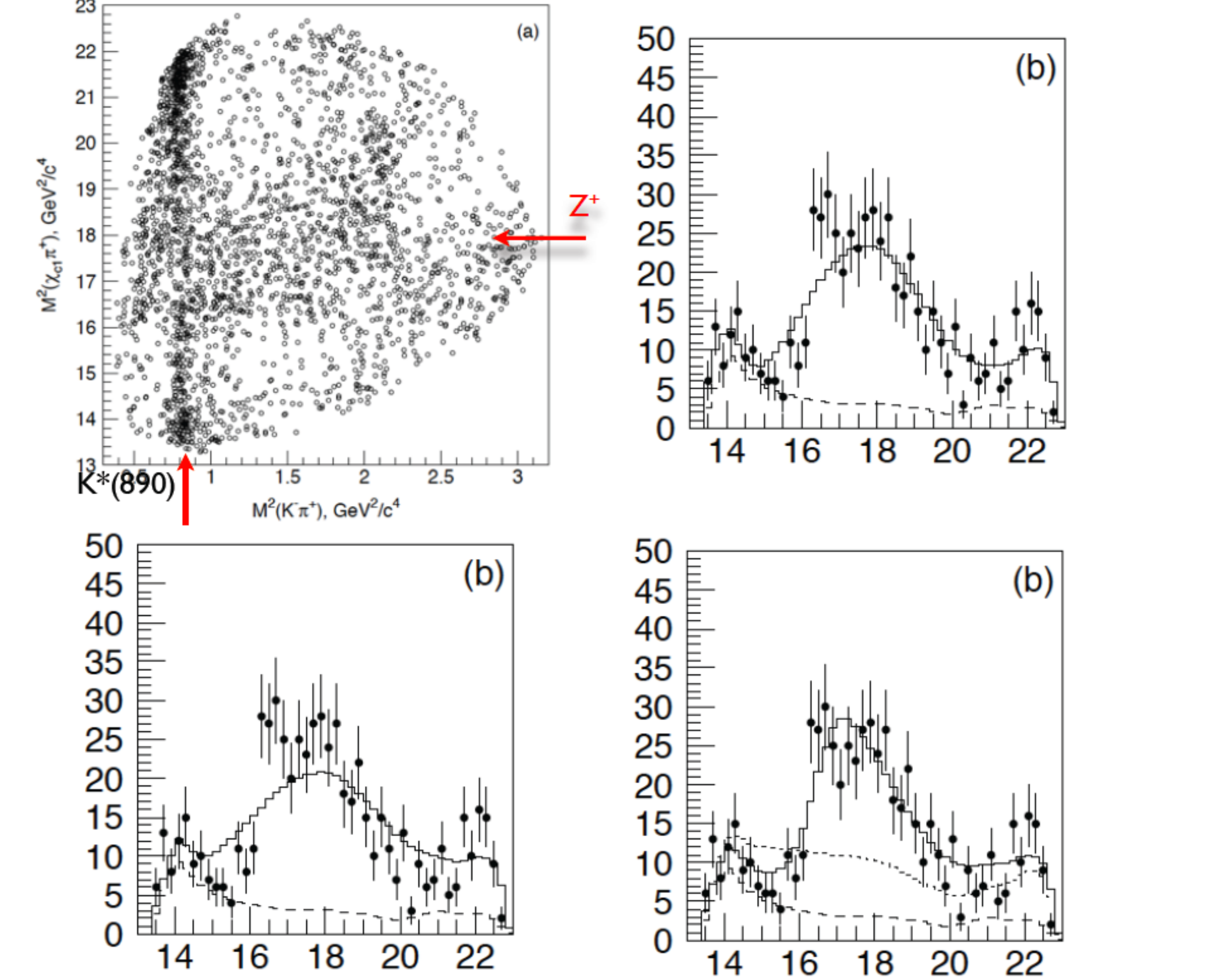}
\caption{Top-Left:  $\chi_{c1}\pi^+ K^-$ Dalitz distribution. The vertical arrow indicates the $K^*$ resonance band  and the horizontal arrow points to the region of possible charged charmonium resonance, $Z^+$. The other panels show the  $\chi_{c1}\pi^+$ mass distribution together with  fit results including: no $Z^+$ (bottom-left), no $Z^+$ but the $K^*_2$ resonance (top-right), single $Z^+$ and no spin-2 $K^*_2$'s (bottom-left). Dashed  line indicates strength of the coherent background. } 
 \label{fig3}
\end{figure}
 
 Similar  analysis was carried out for the 
  $\bar B^0 \to K^- \chi_{c1} \pi^+$ decay. The Dalitz plot is shown in Fig.~\ref{fig4}. In this case 
 enhancement in the horizontal band {\it i.e.} in the $\chi_{c1} \pi^+$ mass distribution is visible. The results of the fit,  with an amplitude set constructed in analogy to the  $\psi'pi K$ analysis  
 also prefers the charged resonances in the  $\chi_{c1}\pi^+$ channel. 
   Kinematic reflection from the $K^*_2$, however,   seem to be quite strong.

  \section{ Bottomonium  spectrum} 
 The $b\bar b$ spectrum bellow the $B\bar B$ decay threshold also follows closely the quark model template. In recent years the Babar~\cite{2011zp} and Belle~\cite{Adachi:2011ji}  
  reported discoveries of  the $h_b(1P)$ and the $h_b(2P)$  states. The lightest bottomonium 
   state, the $\eta_b$,  was  recently  discovered by Babar \cite{:2009pz}
  in the dacay $\Upsilon(2S) \to \gamma \eta_b(1S)$.

\subsection{ Charged bottomonia ?}  
 
In addition to the charged charmonia, Bell searched for charged  bottomonia in the 
Dalitz distribution of the  decays of the $\Upsilon(5S)$ 
 to  $\Upsilon(1S), (2S), (3S)$   accompanied by the $\pi^+\pi^-$ pair \cite{Collaboration:2011gja}. 
  In all cases visible enhancement in the $\Upsilon(nS) \pi$  mass in a band around $M\sim 10.6\mbox{ GeV}$ was observed. The analysis included isoscalar resonances in the  $\pi\pi$ channel   and a smooth non-resonant background and a resonance in the $\Upsilon(nS)\pi$ channel. The analysis details were  discussed  at this conference Ref.~\cite{asner-this-confernce}.

\section{Hybrid mesons} 
\label{hyb} 

\subsection{Theoretical expectations} 
 Hybrid mesons are resonances that originate from excitation of the gluon cloud surrounding  the valence quarks. These gluonic excitations may be associated with those of the flux tube,  however  since in "real" hadrons quarks are either relativistic (light mesons) or close together (heavy  quarkonia)  flux tube excitations may be quite  complicated compared. 
   The additional, gluonic degrees of freedom lead to hybrid mesons with 
    $J^{PC}$ quantum numbers that are not accessible to a valence $Q{\bar Q}$ state. These are referred to as exotic mesons\cite{Meyer:2010ku}.  Recent lattice computations of the 
    meson spectrum,  both for heavy and light quarks 
confirm the quark model nature of the low lying multiplets and indicate existence of exotic mesons, with the lightest exotic having the $J^{PC} = 1^{-+}$ quantum numbers~\cite{Jozef:2010,dud2,other}. In these lattice computations the quark/gluon content was determined   
  by analysis of matrix elements of operators with different combinations of quark and gluon fields. In the $J^{PC}=1^{--}$, vector sector, for example it was found that the first four lightest states (for all flavors)  could  be identified  with  the three  $1S$, $2S$, $1D$ quark model states and  a single hybrid meson. This gives credence to the hypothesis that the $Y(4260)$ is a $c{\bar c}$ hybrid. Lattice simulations also seem to indicate existence of multiplets, beyond those predicted by the quark model that  combine states with a significant  gluonic  content~\cite{joe-latest}.
 In particular the lowest multiplet of hybrid mesons may contain $1^{--}, 0^{-+},1^{-+},2^{-+}$ states 
 that includes the exotic, $1^{-+}$ meson. 
A possible second multiplet   containing, $(0,1,1,1,2,2,3)^{+-}$ and $(0,1,2)^{++}$ could also
 be identified. Appearance of such multiples was predicted in ~\cite{peng1,peng2} based on the analysis of the gluon spectrum emerging from canonical quantization in the Coulomb gauge. For heavy, non-relativistic quarks the Coulomb gauge picture is consistent with the lattice determination of the excited gluon spectrum in presence of static  quark sources. In particular, as the relative separation between the quark and the antiquark vanishes, the exited gluon field forms a glue-lump with the lowest energy states  having quantum quantum numbers $J^{PC}_{glue} = 1^{+-},1^{--},\cdots$~\cite{balli} . Such unusual ordering can be explained by assuming the glue-lump behaves as a quasi particle with natural $1^{--}$ quantum numbers, but the interaction with a color octet, $Q{\bar Q}$  source is spin-dependent and lowers the  energy of  the glue-lump in the $L=1$, $P$-wave state  compared to the $L=0$ $S$-wave state. Two multiplets 
   of hybrid mesons, in agreement with lattice simulations, follow from combining quantum numbers of the glue-lump with those of the $Q{\bar Q}$ pair. 

\subsection{Experimental Status} 
\begin{figure}[ht]
\centering
\includegraphics[width=150mm]{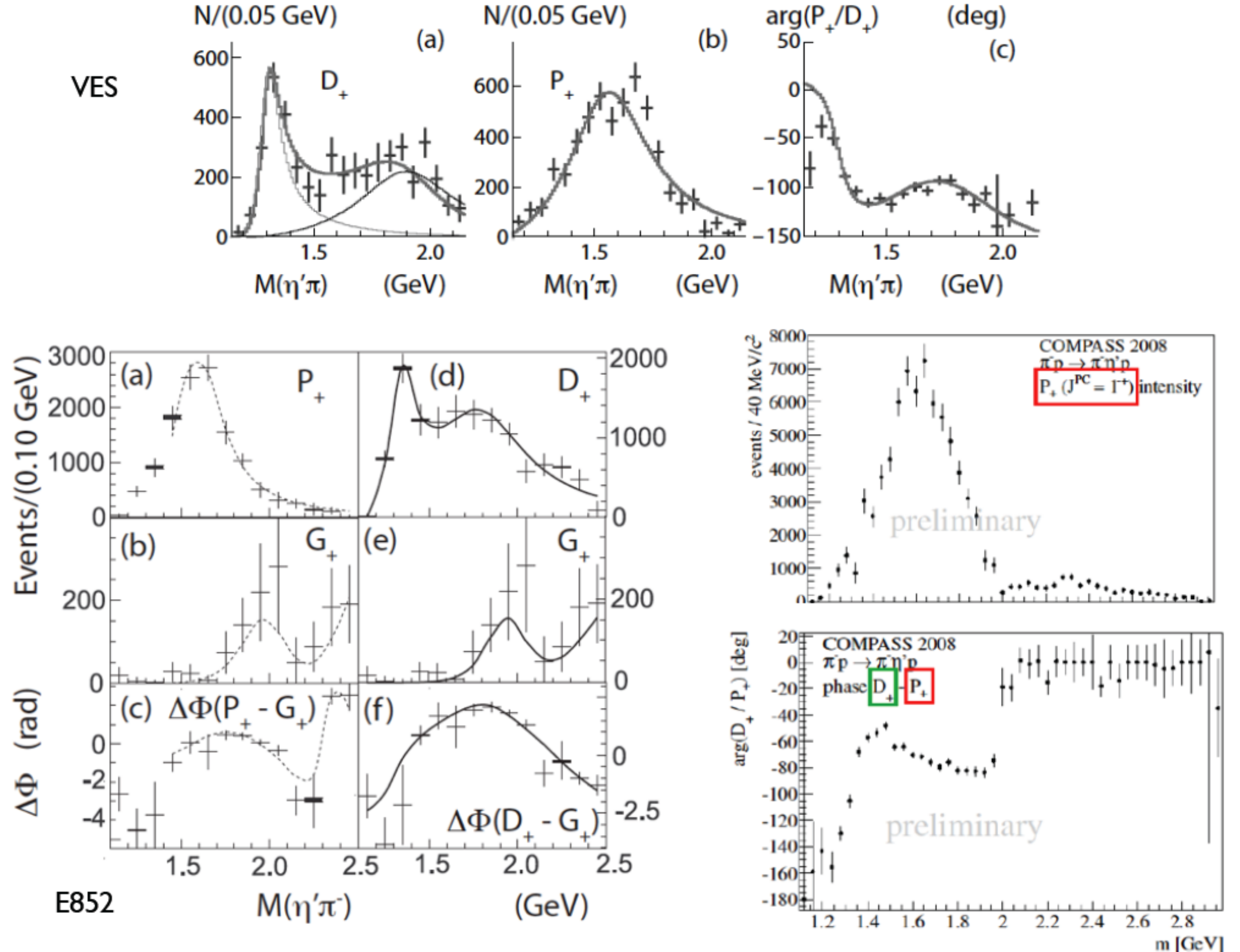}
\caption{ Dominant partial waves in the $\eta' \pi$ system measured by VES (top), E852 (bottom-left), COMPASS (bottom-right). } 
 \label{fig4}
\end{figure}

As discussed earlier the $Y(4260)$  is a possible candidate for a hybrid with non-exotic quantum numbers in the $c\bar c$ sector. The $s\bar s$ analogue might be the $Y(2175)$ seen by Belle  in $e^+ e^- \to \phi \pi^+\pi^-$~\cite{Ys}. In the $u{\bar u}/d {\bar d}$ sector there have been several candidates reported. A review of all experimental results is given in Ref.~\cite{Meyer:2010ku}.  The most recent sighting   comes from the measurement by COMPASS collaboration~\cite{comp}  of the $\eta'\pi'$ angular distribution as a function of  the pair's invariant mass in the reaction $\pi^- p \to \eta'\pi^- p$. The $\eta^{(')} \pi$ system in the relative $P$-wave has exotic quantum numbers $1^{-+}$ thus a resonance in this system could be a candidate for an exotic meson. In the past the VES collaboration~\cite{Beladidze:1993km} reported a strong intensity and phase motion characteristic to a broad resonance. The E852 experiment at BNL   \cite{Ivanov:2001rv} also found a likely resonance with parameters consistent with the VES measurement. By fitting the intensity  and phase of the $\eta'\pi$  the $P$-wave 
  to a Breit-Wigner resonance E852 determined the mass and width of the exotic resonance 
   to be  $M=1.597^{+0.045}_{-0.010} \pm 0.010$, $\Gamma = 0.340 \pm 0.50 \pm 0.040$. 
 Comparison of the VES, E852 and COMPASS results of the $P$-wave are shown in Fig.~\ref{fig4}.

\subsection{Future plans} 
The  COMAPSS collaboration is currently analyzing high statistics production of various 
  meson channels in $\pi^-$ $p$ diffraction at $190\mbox{ GeV}$ on both proton and  nuclear targets.  Another large data set  soon to be available  for studies of light flavor resonances is currently being collected by the BESIII experiment. 
Two new facilitates are expected in the near future: the energy upgraded Jefferson Lab with two new detectors, GlueX and CLAS12 and the PANDS facility at GSI. 

\section{Outlook} 

Information about the hadron spectrum were collected in many different experiments over the last three decades using different probes and targets. Many leading laboratories in the world (BES, CERN, JLab, SLAC) and the new experimental facilities that will be built in a near future (GSI, JLAB12, JPARC) Êdevote an important part of their physics program to light hadron spectroscopy.  Several of these experiments aim at discovery of exotic hadrons. Hybrid mesons, for example  underline the role of gluon self-interactions in hadrons, and provide the basis for understanding the  confinement mechanism that eternally locks quarks and gluons in hadrons.

Another important goal of future hadron physics will be in the precision front. In order to find signatures of physics beyond the standard model it is important to be in the position to analyze hadronic observables with a high and controlled accuracy.  In this context CP violating contributions to the many particle decays of heavy flavors  (i.e. B and D mesons) are of very high interest.

These new experiments with unprecedented statistics, with excellent resolution and with polarization observables measured, demand an unrivaled degree of detail in modeling the dynamics of strong interaction processes if new discoveries and insights into the hadron spectrum are to result. Thus the  success of the new generation of experiments studying the fundamental spectroscopy of hadrons and fundamental symmetries with hadronic reactions depends critically on the robust information that can be extracted about the scattering matrix elements that form the basis of each reaction process.
 
 It is vital that theorists concerned with the physics of reaction mechanisms work with experimentalists to understand the relevant energy-dependent physics backgrounds that could mimic potential signals of a resonant states and that they find a way to extract robust information from data about scattering matrix elements.
 
\begin{acknowledgments}
This work was supported in part by US DOE under contract DE-FG0287ER40365
\end{acknowledgments}

\bigskip % extra skip inserted

\begin{thebibliography}{99}   % Use for  1-9  references
%\begin{thebibliography}{99} % Use for 10-99 references



%\cite{Collins:1971ff}
\bibitem{Collins:1971ff}
  P.~D.~B.~Collins,
  %``Regge theory and particle physics,''
  Phys.\ Rept.\  {\bf 1}, 103 (1971).
  


  \bibitem{Jozef:2010}
J. J. Dudek, R. G. Edwards, M. J. Peardon, D. G. Richards, and C. E. Thomas, 
  Phys.\ Rev.\  D. {\bf 82}, 034508(2010).
  %% Toward the excited meson spectrum of dynamical QCD.

\bibitem{Pelaez:2003dy}
  J.~R.~Pelaez,
  %``On the Nature of light scalar mesons from their large N(c) behavior,''
  Phys.\ Rev.\ Lett.\  {\bf 92}, 102001 (2004)
  [arXiv:hep-ph/0309292].
  



  \bibitem{Martin:1978}
PA. D. Martin and M. R. Pennington,
  Annals.\ Phys.\  {\bf 114}, 1(1978).
  %% How imposing analyticity on a ¹¹ phase shift analysis can reveal new solutions, explore experimental structures, and investigate the possibility of new resonances.

%\cite{Greensite:2003bk}
\bibitem{Greensite:2003bk}
  J.~Greensite,
  %``The Confinement problem in lattice gauge theory,''
  Prog.\ Part.\ Nucl.\ Phys.\  {\bf 51}, 1 (2003)
  [arXiv:hep-lat/0301023].
  %%CITATION = PPNPD,51,1;%%


\bibitem{pdg} K. Nakamura et al. (Particle Data Group), J. Phys. G 37, 075021 (2010).  


%\cite{Choi:2003ue}
\bibitem{Choi:2003ue}
  S.~K.~Choi {\it et al.}  [Belle Collaboration],
  %``Observation of a narrow charmonium - like state in exclusive B+- ---> K+-
  %pi+ pi- J / psi decays,''
  Phys.\ Rev.\ Lett.\  {\bf 91}, 262001 (2003)
  [arXiv:hep-ex/0309032].
  %%CITATION = PRLTA,91,262001;%%
  
  %\cite{Acosta:2003zx}
\bibitem{Acosta:2003zx}
  D.~Acosta {\it et al.}  [CDF II Collaboration],
  %``Observation of the narrow state $X(3872) \to J/\psi \pi^+ \pi^-$ in
  %$\bar{p}p$ collisions at $\sqrt{s} = 1.96$ TeV,''
  Phys.\ Rev.\ Lett.\  {\bf 93}, 072001 (2004)
  [arXiv:hep-ex/0312021].
  %%CITATION = PRLTA,93,072001;%%
  
  
  %\cite{Swanson:2006st}
\bibitem{Swanson:2006st}
  E.~S.~Swanson,
  %``The New heavy mesons: A Status report,''
  Phys.\ Rept.\  {\bf 429}, 243 (2006)
  [arXiv:hep-ph/0601110].
  %%CITATION = PRPLC,429,243;%%
  
  \bibitem{jo} 
  P.del Amo Sanchez {\it et al.} [BABAR Collaboration], Phys. Rev. D {\bf 82}, 132002 (2010). 
  
  %\cite{Suzuki:2005ha}
\bibitem{Suzuki:2005ha}
  M.~Suzuki,
  %``The X(3872) boson: Molecule or charmonium,''
  Phys.\ Rev.\  D {\bf 72}, 114013 (2005)
  [arXiv:hep-ph/0508258].
  %%CITATION = PHRVA,D72,114013;%%

%\cite{Bhardwaj:2011dj}
\bibitem{Bhardwaj:2011dj}
  V.~Bhardwaj {\it et al.}  [Belle Collaboration],
  %``Observation of $X(3872)\to J/\psi \gamma$ and search for
  %$X(3872)\to\psi'\gamma$ in B decays,''
  Phys.\ Rev.\ Lett.\  {\bf 107}, 9 (2011)
  [arXiv:1105.0177 [hep-ex]].
  %%CITATION = PRLTA,107,9;%%
  
  
\bibitem{pelaez-kaminski}
  R.~Garcia-Martin, R.~Kaminski, J.~R.~Pelaez and J.~Ruiz de Elvira,
  %``Precise determination of the f0(600) and f0(980) pole parameters from a
  %dispersive data analysis,''
  Phys.\ Rev.\ Lett.\  {\bf 107}, 072001 (2011)
  [arXiv:1107.1635 [hep-ph]].

  
  %\cite{Aubert:2005rm}
\bibitem{Aubert:2005rm}
  B.~Aubert {\it et al.}  [BABAR Collaboration],
  %``Observation of a broad structure in the $\pi^+ \pi^- J/\psi$ mass spectrum
  %around 4.26-GeV/c$^2$,''
  Phys.\ Rev.\ Lett.\  {\bf 95}, 142001 (2005)
  [arXiv:hep-ex/0506081].
  %%CITATION = PRLTA,95,142001;%%
  
  
  %\cite{delAmoSanchez:2010yp}
\bibitem{delAmoSanchez:2010yp}
  P.~del Amo Sanchez {\it et al.}  [The BABAR Collaboration],
  %``Dalitz plot analysis of $D_s^+ \to K^+ K^- \pi^+$,''
  Phys.\ Rev.\  D {\bf 83}, 052001 (2011)
  [arXiv:1011.4190 [hep-ex]].
  %%CITATION = PHRVA,D83,052001;%%
  
  
%  \cite{:2007sj}
\bibitem{:2007sj}
  C.~Z.~Yuan {\it et al.}  [Belle Collaboration],
  %``Measurement of e+ e- ---> pi+ pi- J/psi cross-section via initial state
  %radiation at Belle,''
  Phys.\ Rev.\ Lett.\  {\bf 99}, 182004 (2007)
  [arXiv:0707.2541 [hep-ex]].
  %%CITATION = PRLTA,99,182004;%%
  
  
  %\cite{He:2006kg}
\bibitem{He:2006kg}
  Q.~He {\it et al.}  [CLEO Collaboration],
  %``Confirmation of the Y(4260) resonance production in ISR,''
  Phys.\ Rev.\  D {\bf 74}, 091104 (2006)
  [arXiv:hep-ex/0611021].
  %%CITATION = PHRVA,D74,091104;%%


\bibitem{felipe} F.J.Llanes-Estrada, Phys. Rev. D {|bf 72}, 031503 (2005). 


%\cite{Abe:2004zs}
\bibitem{Abe:2004zs}
  K.~Abe {\it et al.}  [Belle Collaboration],
  %``Observation of a near-threshold omega J/psi mass enhancement in exclusive B
  %---> K omega J/psi decays,''
  Phys.\ Rev.\ Lett.\  {\bf 94}, 182002 (2005)
  [arXiv:hep-ex/0408126].
  %%CITATION = PRLTA,94,182002;%%
  
  
  %\cite{Abe:2007jn}
\bibitem{Abe:2007jn}
  K.~Abe {\it et al.}  [Belle Collaboration],
  %``Observation of a new charmonium state in double charmonium production in e+
  %e- annihilation at s**(1/2) ~ 10.6-GeV,''
  Phys.\ Rev.\ Lett.\  {\bf 98}, 082001 (2007)
  [arXiv:hep-ex/0507019].
  %%CITATION = PRLTA,98,082001;%%

  
  
  %\cite{delAmoSanchez:2010jr}
\bibitem{delAmoSanchez:2010jr}
  P.~del Amo Sanchez {\it et al.}  [BABAR Collaboration],
  %``Evidence for the decay X(3872) ---> J/ psi omega,''
  Phys.\ Rev.\  D {\bf 82}, 011101 (2010)
  [arXiv:1005.5190 [hep-ex]].
  %%CITATION = PHRVA,D82,011101;%%
  
  
  %\cite{Aaltonen:2011at}
\bibitem{Aaltonen:2011at}
  T.~Aaltonen {\it et al.}  [The CDF Collaboration],
  %``Observation of the $Y(4140)$ structure in the $J/\psi\,\phi$ Mass Spectrum
  %in $B^\pm\to J/\psi\,\phi K$ decays,''
  arXiv:1101.6058 [hep-ex].
  
  

%\cite{:2009da}
\bibitem{:2009da}
  R.~Mizuk {\it et al.}  [BELLE Collaboration],
  %``Dalitz analysis of B ---> K pi+ psi-prime decays and the Z(4430)+,''
  Phys.\ Rev.\  D {\bf 80}, 031104 (2009)
  [arXiv:0905.2869 [hep-ex]].
  %%CITATION = PHRVA,D80,031104;%%

  
  %\cite{Dzierba:2003cm}
\bibitem{Dzierba:2003cm}
  A.~R.~Dzierba, D.~Krop, M.~Swat, S.~Teige and A.~P.~Szczepaniak,
  %``The Evidence for a pentaquark signal and kinematic reflections,''
  Phys.\ Rev.\  D {\bf 69}, 051901 (2004)
  [arXiv:hep-ph/0311125].
  %%CITATION = PHRVA,D69,051901;%%
  
  
%\cite{:2008nk}
\bibitem{:2008nk}
  B.~Aubert {\it et al.}  [BABAR Collaboration],
  %``Search for the Z(4430)- at BABAR,''
  Phys.\ Rev.\  D {\bf 79}, 112001 (2009)
  [arXiv:0811.0564 [hep-ex]].
  %%CITATION = PHRVA,D79,112001;%%




%\cite{Mizuk:2008me}
\bibitem{Mizuk:2008me}
  R.~Mizuk {\it et al.}  [Belle Collaboration],
  %``Observation of two resonance-like structures in the pi+ chi(c1) mass
  %distribution in exclusive anti-B0 ---> K- pi+ chi(c1) decays,''
  Phys.\ Rev.\  D {\bf 78}, 072004 (2008)
  [arXiv:0806.4098 [hep-ex]].
  %%CITATION = PHRVA,D78,072004;%%

  
  %\cite{:2011zp}
\bibitem{2011zp}
   and J.~P.~Lees  [The BABAR Collaboration],
  %``Evidence for the h_b(1P) meson in the decay Upsilon(3S) --> pi0 h_b(1P),''
  arXiv:1102.4565 [hep-ex].
  %%CITATION = ARXIV:1102.4565;%%
  
  
  %\cite{Adachi:2011ji}
\bibitem{Adachi:2011ji}
  I.~Adachi {\it et al.}  [Belle Collaboration],
  %``First observation of the P-wave spin-singlet bottomonium states h_b(1P) and
  %h_b(2P),''
  arXiv:1103.3419 [hep-ex].
  %%CITATION = ARXIV:1103.3419;%%

  
  
  %\cite{:2009pz}
\bibitem{:2009pz}
  B.~Aubert {\it et al.}  [BABAR Collaboration],
  %``Evidence for the eta(b)(1S) Meson in Radiative Upsilon(2S) Decay,''
  Phys.\ Rev.\ Lett.\  {\bf 103}, 161801 (2009)
  [arXiv:0903.1124 [hep-ex]].
  %%CITATION = PRLTA,103,161801;%%



%\cite{Collaboration:2011gja}
\bibitem{Collaboration:2011gja}
  B.~Collaboration,
  %``Observation of two charged bottomonium-like resonances,''
  arXiv:1105.4583 [hep-ex].
  %%CITATION = ARXIV:1105.4583;%%

\bibitem{asner-this-confernce} 
D. Asner [Belle Collaboration], {\it at this conference} 



\bibitem{dud2} J.J.Dudek {\it et al.} [Hadron Spectrum Collaboration], Phys. Rev. D{\bf 77} 034501 (2008). 



   
   %\cite{Meyer:2010ku}
\bibitem{Meyer:2010ku}
  C.~A.~Meyer and Y.~Van Haarlem,
  %``The Status of Exotic-quantum-number Mesons,''
  Phys.\ Rev.\  C {\bf 82}, 025208 (2010)
  [arXiv:1004.5516 [nucl-ex]].
  %%CITATION = PHRVA,C82,025208;%%

\bibitem{other} see Ref.~\cite{Meyer:2010ku} for summary of earlier predictions. 


 \bibitem{joe-latest}
  J.~J.~Dudek,
  %``The lightest hybrid meson supermultiplet in QCD,''
  arXiv:1106.5515 [hep-ph].
  %%CITATION = ARXIV:1106.5515;%%


\bibitem{peng1} 
  P.~Guo, A.~P.~Szczepaniak, G.~Galata, A.~Vassallo and E.~Santopinto,
  %``Heavy quarkonium hybrids from Coulomb gauge QCD,''
  Phys.\ Rev.\  D {\bf 78}, 056003 (2008)
  [arXiv:0807.2721 [hep-ph]].
  %%CITATION = PHRVA,D78,056003;%%

\bibitem{peng2}  P.~Guo, A.~P.~Szczepaniak, G.~Galata, A.~Vassallo and E.~Santopinto,
  %``Gluelump spectrum from Coulomb gauge QCD,''
  Phys.\ Rev.\  D {\bf 77}, 056005 (2008)
  [arXiv:0707.3156 [hep-ph]].
  %%CITATION = PHRVA,D77,056005;%%


\bibitem{balli} 
  G.~S.~Bali and A.~Pineda,
  %``QCD phenomenology of static sources and gluonic excitations at short
  %distances,''
  Phys.\ Rev.\  D {\bf 69}, 094001 (2004)
  [arXiv:hep-ph/0310130].
  %%CITATION = PHRVA,D69,094001;%%



\bibitem{Ys} 
C.P.~Shen {\it et al.}  [Belle Collaboration],
  %``First observation of the P-wave spin-singlet bottomonium states h_b(1P) and
  %h_b(2P),''
 Phys. Rev. D {\bf 80}, 031101(R) (2009) 
   %%CITATION = ARXIV:1103.3419;%%
   

\bibitem{comp} 
T.Schl\"uter, at  Hadron 2011, Munich 2011.



   %\cite{Beladidze:1993km}
\bibitem{Beladidze:1993km}
  G.~M.~Beladidze {\it et al.}  [VES Collaboration],
  %``Study of pi- N ---> eta pi- N and pi- N ---> eta-prime pi- N reactions at
  %37-GeV/c,''
  Phys.\ Lett.\  B {\bf 313}, 276 (1993).
  %%CITATION = PHLTA,B313,276;%%
   
   
  % \cite{Ivanov:2001rv}
\bibitem{Ivanov:2001rv}
  E.~I.~Ivanov {\it et al.}  [E852 Collaboration],
  %``Observation of exotic meson production in the reaction pi- p ---> eta-prime
  %pi- p at 18-GeV / c,''
  Phys.\ Rev.\ Lett.\  {\bf 86}, 3977 (2001)
  [arXiv:hep-ex/0101058].
  %%CITATION = PRLTA,86,3977;%%
   
   
\end{thebibliography}

\end{document}